\documentclass[usenatbib]{mn2e}

\usepackage{graphicx}
\usepackage{natbib}
\def\citeN{\citet}
\def\cite{\citep}

\footnotesize
\newdimen\digitwidth    
\setbox0=\hbox{\rm0}
\digitwidth=\wd0
\catcode`!=\active
\def!{\kern\digitwidth}
\normalsize
\title[Discovery of 28 pulsars]{Discovery of 28 pulsars using new techniques for sorting pulsar candidates}
\author[M.~J.~Keith et al.]
{M.~J.~Keith$^{1,2}$\thanks{Email: mkeith@pulsarastronomy.net},
R.~P.~Eatough$^1$,
A.~G.~Lyne$^1$,
M.~Kramer$^1$,
A.~Possenti$^3$,\newauthor
F.~Camilo$^4$ and
R.~N.~Manchester$^2$
\\
$^1$ University of Manchester, Jodrell Bank Centre for Astrophysics, Alan Turing Building, Manchester M13 9PL, UK\\
$^2$ Australia Telescope National Facility, CSIRO, P.O. Box 76, Epping, NSW 1710, Australia\\
$^3$ INAF - Osservatorio Astronomico di Cagliari, Poggio dei Pini, 09012 Capoterra, Italy\\
$^4$ Columbia Astrophysics Laboratory, Columbia University, New York, NY 10027, USA\\
}
\date{}
\begin{document}

\maketitle
\newcommand{\setthebls}{
}

\setthebls

\begin{abstract} 
Modern pulsar surveys produce many millions of candidate pulsars, far
more than can be individually inspected. Traditional methods for filtering
these candidates, based upon the signal-to-noise ratio of the
detection, cannot easily distinguish between interference signals and
pulsars. We have developed a new method of scoring candidates using a
series of heuristics which test for pulsar-like properties of the
signal. This significantly increases the sensitivity to weak pulsars
and pulsars with periods close to interference signals. By applying
this and other techniques for ranking candidates from a previous
processing of the Parkes Multi-beam Pulsar Survey, 28 previously
unknown pulsars have been discovered.  These include an eccentric
binary system and a young pulsar which is spatially coincident with a
known supernova remnant.

\end{abstract}


\section{Background}
\label{background}
The Parkes Multi-beam Pulsar Survey (PMPS) is the most successful
large-scale pulsar survey ever undertaken.  Processing and
re-processing of the PMPS data set has led to the discovery and
publication of more than 750 previously unknown pulsars.  A full
description of the survey and the discoveries to date are outlined in
the following papers. \citeN{mlc+01} discussed the hardware and
software configuration for the survey and presented the first 100 new
pulsars.  \citeN{mhl+02} provided details of a further 120 pulsars,
and performed an initial statistical analysis of the the
discovered pulsars.  Two hundred more discoveries, including young pulsars and
their association with EGRET sources are discussed by \citeN{kbm+03}.
\citeN{hfs+04} presented a further 180 discoveries and discussed the 
detection of 281 previously-known pulsars in the survey.
A complete re-processing of the survey data using new search
algorithms was presented by \citeN{fsk+04}, hereafter known as the
2002 re-processing.  \citeN{lfl+06} presented 142 discoveries
resulting from that analysis and performed a detailed statistical
study of the pulsar population in the survey region.  Most recently
\citeN{mll+06} detected 11 new transient radio pulsars, the
RRATs which are only detected through their single pulses.

Section \ref{jreaper} describes new techniques for selecting candidate
pulsars for re-observation from the many millions of candidates that
are produced in the processing of a large-scale pulsar survey.  In Section
\ref{discoveries} we describe how we applied these techniques to the candidates from the
2002 re-processing and produced a short-list of 44 new candidates.
These candidates were re-observed and
yielded 28 previously unknown pulsars.  Full astrometric and spin
parameters for each of the newly discovered pulsars have been measured
and are included in Section \ref{details}.  Additionally we discuss
two of the notable discoveries in detail.  To determine the
effectiveness of the new selection techniques, Section \ref{stats}
includes a statistical comparison of the new discoveries and those
presented by \citeN{fsk+04}.

\section{Candidate selection}
\label{jreaper}
In a typical pulsar survey hundreds of pulsar candidates are generated by the
search software for each beam in the survey. With large surveys, this can add
up to millions of candidate pulsars.
Each candidate consists of a selection of candidate parameters, which include the optimised period and dispersion measure (DM) of the pulsar as well as
the signal-to-noise ratio as detected in the various search methods.
Additionally, an number of diagnostic plots are generated from the time-series data, including a folded profile of the pulsar, the variation of signal-to-noise ratio with
DM and plots showing the variation of pulse profile with time and observing frequency.
These parameters and plots are used to select the candidate pulsars that are to be re-observed.

\begin{figure}
\centering
\includegraphics[width=7.5cm]{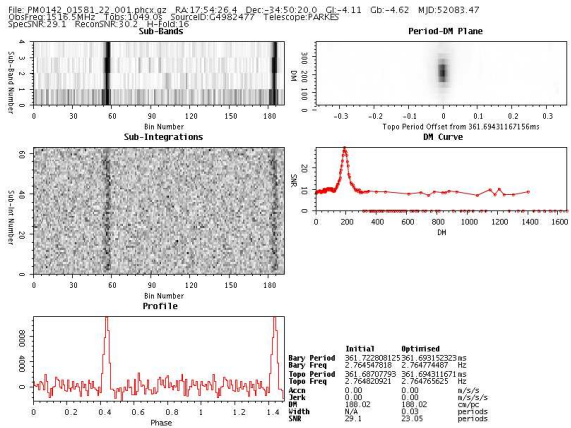}
\caption[A example display showing a single pulsar candidate]{
\label{example_plot}
An example display of the data generated for a single candidate produced in a pulsar search.
This plot shows the detection of the known pulsar PSR~J1754--3443.
The three plots on the left hand side are: the sub-bands plot, showing the four stacked profiles at different observing frequencies;
the sub-integrations plot, indicating how the folded pulse profile is varying with time;
and the pulse profile, folded at the optimal period and DM.
On the right of the figure are two plots: The period-DM diagram, which shows the variation of signal-to-noise ratio with small changes in folding period; and DM and
the DM curve, which shows the variation of signal-to-noise ratio over a large range of DM values.
Also included are the pre and post-optimisation parameters and some basic header information about the observation.
}
\end{figure}

During the 2002 re-processing of the PMPS, candidates were selected
for re-observation using {\sc Reaper}, a graphical selection tool
described by \citeN{fsk+04}.  In brief, {\sc Reaper} presents each
candidate pulsar as a single point on a phase-space diagram.  This
diagram shows many thousands of candidates as a two-dimensional graph showing, for example, period on the
x-axis and signal-to-noise ratio on the y-axis.  This aids the
selection of candidate points that are sufficiently distinguished from
noise and interference signals.  Selection of a point on this display
causes the details of the selected candidate to be displayed to the
user who can then make a decision regarding whether the candidate
should be re-observed.
An example of the plots that are presented when viewing the full candidate
details are shown in Figure \ref{example_plot}.

\begin{figure}
\centering
\includegraphics[width=6cm,angle=-90]{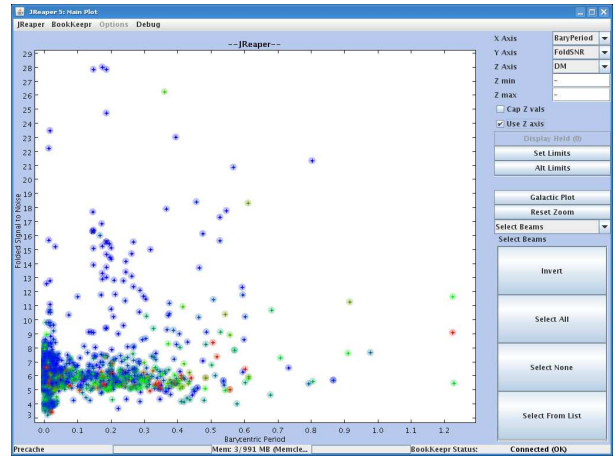}
\caption[Screen-grab of a typical {\sc JReaper} session]{
\label{jreaper_grab}
A typical {\sc JReaper} display showing a number of candidates plotted with barycentric period verses signal-to-noise ratio.
Each point is a pulsar candidate, and is coloured according to a third parameter (in this case the peak DM).
Clicking on a candidate brings up a window showing full details of the candidate, including the diagnostic plots, and gives options to mark the candidate as a potential target for follow-up observations.
}
\end{figure}

Following on the success of {\sc Reaper}, a new software tool, {\sc
JReaper}\footnote{{\sc JReaper} was originally so named as it was a Java implementation of {\sc Reaper}}, has been developed.  This new tool has been developed to
provide the functionality of {\sc Reaper}, as well as advanced new
functionality, on a wider range of data formats.  This has been
achieved by using a modular design which uses generic interfaces to
read and write many different data formats.  This allows for {\sc
JReaper} to be used on many different surveys and data archives
without any significant changes to the core code base. 
Figure \ref{jreaper_grab} shows a screen grab from an example session with {\sc JReaper}, in this case showing candidate pulsars on a plot of barycentric period and folded signal-to-noise ratio.

{\sc JReaper} has a number of features that have been developed to help pulsar searching.
These include many simple functions, such as tracking viewed candidates and filtering displayed candidates by user selectable parameters (for example, hiding candidates with a DM of less than some value).
The rest of this section provides details on some of the new, more advanced, functions of {\sc JReaper}.

\subsection{Scoring}
\label{scoring}
Whilst the human-eye aspects of pulsar searching cannot easily be
replicated by a machine, it is possible to detect some of the basic
indicators used by human observers to distinguish real pulsars from
 interference.  {\sc JReaper} features a scoring
algorithm that attempts to do this by assigning a numerical score to
each candidate pulsar signal based on the data that are produced by the search
algorithms.
The scoring routine uses a number of tests that are then combined with
weights to give a final score.  The tests currently used are as
follows:

\subsubsection{Sub-integration scoring}
Typical pulsars have a period and profile shape that are completely stable
over the observation time, whereas for periodic man-made signals the period is often more variable.
Periodic signals can be identified by eye by looking for straight lines
running down a plot of pulse phase vs observation time, known as the
sub-integration plot.  We have investigated a number of ways to
automate this process.
\begin{enumerate}

\item 2-Dimensional Kolmogorov-Smirnov Test -- The Kolmogorov-Smirnov
test provides a statistical way to determine if two data sets differ
from one another significantly. This can be applied to the
sub-integrations by comparing the observed values with a sample set
generated from Gaussian noise. Any significant deviation implies that
the signal is not due simply to noise. Whilst effective at
discriminating between
signals and noise, this test is not effective at differentiating
between interference signals and real pulsar signals. It is also
ineffective when the pulse is weak enough that there is not a
significant signal in each sub-integration. For these reasons the
Kolmogorov-Smirnov Test was not used in the final version of {\sc
JReaper}.
\item Hough Transform -- Under the Hough transform, the input data is
mapped in terms of any set of parametrised curves
\cite{hou60,dh72}. For the purpose of finding straight lines this
means showing the data as a plot of how much power lies on each line
with a given slope and intercept.
This can be used to simplify the
problem of finding a straight line down the sub-integrations to that
of finding a discrete significant peak in the Hough transformed
plane. In this case, each slope corresponds to an error in the folding period and 
intercept to a phase bin of the profile.
This can be generalised to any polynomial to fit accelerated
sources, where the pulse phase drifts non-linearly with time, although the
dimensionality of the Hough plane increases with each extra
power. This method is effective at removing signals that are not
caused by strictly periodic sources, and is not dependant on the
strength of the signal in each sub-integration. However the Hough
transform is computationally expensive and is effectively repeating
work that was already done as part of the time domain
optimisation. The Hough transform is computationally
equivalent to the period adjustment performed by sliding and adding
the sub-integrations.
This operation is performed during the final `time-domain optimisation'
step of typical pulsar search methods.
Therefore with simple modifications to existing code, the Hough method
can be applied to data without any extra processing cost.

\item Period curve scoring -- This method considers the variation of
signal-to-noise ratio with trial period as a marker of the periodicity
in the sub-integrations. This works using a similar concept to the
Hough transform, except the phase information is removed by computing
the signal-to-noise ratio of each output profile. This one dimensional
output is much simpler to score, as one can model the expected curve
for a realistic source by:

\begin{equation}
W_{\rm eff} = \sqrt{W_{\rm int}^2 + (T_{\rm obs} \cdot \Delta{P}/P)^2 },
\end{equation}\begin{equation}
\label{broadendsnr}
\frac{\rm Signal}{\rm Noise} \propto \sqrt{P - W_{\rm eff}\over W_{\rm eff}}.
\end{equation}
Here $P$ is the pulse period, $\Delta{P}$ is the offset in period,
$W_{\rm int}$ is the intrinsic pulse width, assumed to be the pulse width
at the optimised period. Note that if $W_{\rm eff}$ is greater than the
period, it is assumed that the pulse is completely smeared, and the
signal-to-noise ratio is 0.

By computing the RMS deviation from the model curve, the observed
curve can be scored, with low RMS scoring highly. This is the
preferred method for scoring the sub-integrations as it a simple
calculation and does not rely on any complex 2-dimensional analysis.

\end{enumerate}
\subsubsection{DM curve scoring}
This scoring method considers how the detected signal-to-noise ratio
varies with trial DM.  Given a period, pulse width and
observational frequency and bandwidth, it is possible to compute a
theoretical variation of signal-to-noise ratio with DM, similar to
that described for period offset above:
\begin{equation}
W_{\rm eff} = \sqrt{W_{\rm int}^2 + (k_{\rm DM} \cdot \Delta{DM} \cdot \Delta{\nu}/\nu^3 )^2 }.
\end{equation}
Here $\Delta{DM}$ is the DM offset from the true DM (measured in cm$^{-3}$pc), $\Delta{\nu}$ is
the total observation bandwidth (in MHz), $\nu$ is the observation frequency (also in MHz) and the
constant $k_{\rm DM} = 8.297616\times10^{3}$ cm$^{3}$pc$^{-1}$s$^{-1}$.  $W_{\rm eff}$ can then be fed into
equation \ref{broadendsnr} above.  This model is then compared with the
observed values by taking the RMS deviation from the theoretical
values.  To avoid bad scores from noise being detected at the same
period, the DM-curve is only considered over the range of DM values for which 
the signal-to-noise ratio less than 20\% of the peak
signal-to-noise ratio.  Candidates score highly if their
RMS deviation is low.

\subsubsection{Zero DM comparison}
Some un-dispersed signals can be mis-detected at a high DM, caused
when a signal with low variation with DM (i.e. a narrow band interference signal) is mixed with a noisy
background. To identify these sources, the DM curve scoring is
re-performed assuming that the source was at DM of zero. If the
scoring is the same or better assuming zero DM then there is a high
probability that the DM value has been overestimated. This routine returns
a high score if the difference between the zero DM score and the
optimised DM score is large.

\subsubsection{Profile scoring}
Profile scoring detects what fraction of the pulse is above a
threshold of 60\% of the peak value.  This is designed to select
against a common type of RFI which exhibits negative `pulses',
 those that have troughs in power rather than peaks.  This scoring method is
tuned to start selecting against profiles which have more than 50\% of
their signal above the threshold.
\subsubsection{Score combination}
The result of each score is stored with each candidate, and the final
score value that is used for plotting is computed based on user
specified weights.  These weights allow the user to display the
results in multiple different views. For example, weighting the sub
integration scoring highly removes a lot of interference, but also
affects binary sources, so the user may choose to view the data twice,
once with high weighting and once with a low weighting to the sub
integration scoring.

Interference signals are often detected with a high signal-to-noise
ratio, although typically they do not show the basic characteristics of
a pulsar and therefore score relatively badly compared to a real
pulsar signal.  The scoring functions are tuned to select signals that
have typical pulsar properties, however this means that it is only
likely to detect typical pulsars.  For example, due to their changing
period over time, binary pulsars often score lower in the
sub-integration scoring than solitary pulsars, although this effect is
mitigated by taking an average of all the tests.

An alternative method for combining scores is to use a neural
network, trained on a large number of real pulsars (including binaries etc.), interference and noise
signals (Eatough et al., in preparation). This can potentially perform far better
than a simple averaging technique.

The advantage of scoring candidates is that it can be used in
conjunction with other candidate selection methods.  Generally, the
more ways of viewing the data, the less likely a pulsar in the data
set will be missed.
\subsection{Harmonic detection}
In surveys covering very pulsar-rich regions of the sky, such as the
PMPS, one of the biggest sources of false positive new pulsar candidate are previously
known pulsars.  These signals are useful for confirming that the
survey is working as expected, although when looking for new pulsars,
re-detections of existing pulsars waste much time and resources.
Whilst it is simple to remove candidates with the period and the position of
known pulsars, they often have many harmonics, including non-integer
fractions of the period.  Unfortunately another real pulsar can easily have
a fundamental period that overlaps with one of these harmonics, so that
removal of all sources that share these periods can possibly remove
potential new pulsars.

To overcome this problem, {\sc JReaper} has an additional check for
harmonics that uses a peak counting method.  This determines the
likelihood of a detection being a given harmonic by comparing the
expected number of peaks with the number that can be counted in the
folded profile of the detection.  For example, a profile with a single
peak cannot be a 1/2 harmonic of a pulsar, similarly a signal with
7/5 of the period of a nearby known pulsar which shows 5 peaks is
likely to indeed be due to that source.  For safety, a comment is made
on any detection that matches a harmonic period, even if it is not
actually marked as a known signal.

\subsection{`3-D' plotting}
As well as the standard two-axis plotting, {\sc JReaper} allows for
colouring of the data points to produce a `3-D' style plot.  This can
be used effectively to distinguish points that are close to each other
in the standard view.  For example, when plotting period against
signal-to-noise ratio, showing detected DM on the
colour axis can help pulsars at higher DM values to be distinguished
from interference with similar periodicity.

\subsection{Plot area selection}
Often a user wishes to view all candidates from a given region of the
{\sc JReaper} screen.  For example, the user has filtered the
candidates and wants to select all the points down to a given
signal-to-noise ratio threshold.  {\sc JReaper} therefore allows the
user to select an area of the screen and sequentially display details for each
of the candidates from that area, excluding any that have been marked as
interference, or have already been viewed.  This allows users to
quickly perform a `blind' search without repetitive inspection of
individual data points.  Using this option, the best candidates can be
identified in a less subjective manner, so reducing the number of
user-based selection effects.

\section{Discovery and timing of 28 Pulsars}
\label{discoveries}

The candidates from the 2002 reprocessing of the PMPS were read into
{\sc JReaper}.  These were then analysed in groups of approximately 10
survey pointings, using multiple {\sc JReaper} views including the new
scoring system described in Section \ref{scoring}.  Good
candidates were classified with a class of 1, 2 or 3, where those classed 1 are
considered most likely to be new pulsars and 3 the least likely.
These short-listed candidates were automatically added to the
processing summary website, which allowed the generation of telescope
commands for confirmation observations.  Such observations were
scheduled initially to observe the class 1 candidates, then to observe
the best from classes 2 and 3.

Using these methods, 44 new candidate pulsars were selected for
re-observation, yielding a total of 28 confirmed new
pulsars.  The confirmation success rate for class 1 was $95\%$, for
class 2 was $64\%$ and for class 3 was $35\%$.  Confirmation of
candidates was carried out by performing a `grid' observation where
five short observations are carried out, one on the target position
and four offset from the target position by half a beam width in each
of the cardinal directions \cite{mhl+02}.  The resulting detections,
or non-detections, and their relative signal strength can provide an
improved position for the pulsar.  If there are no detections in the
gridding observations, a longer survey-length observation is carried
out on the target position.  Typically a non-detection in the longer
observation implies that the candidate is not a real source, although
there may be complications arising from intermittent pulsars
\cite{klo+06} or interstellar scintillation, causing the pulsar flux density to
vary. For these reasons convincing candidates were re-observed several
times, although the possibility that they include very long term nulling
pulsars cannot be ruled out.

Following the procedure established by the timing of pulsars
discovered previously in the PMPS, each of the new pulsars was
regularly observed using the Parkes radio telescope or the Lovell
telescope at Jodrell Bank, for a duration of at least one year. Pulse
time-of-arrival (TOA) measurements were obtained using standard pulsar
timing methodology (e.g. \citealp{lk05}).  These TOAs were then used
to obtain the pulsar's spin, positional and binary
parameters using the {\sc Tempo2} software \cite{hem06}.

\section{The new pulsars}
\label{details}
\begin{figure*}\centering
\includegraphics[width=15cm]{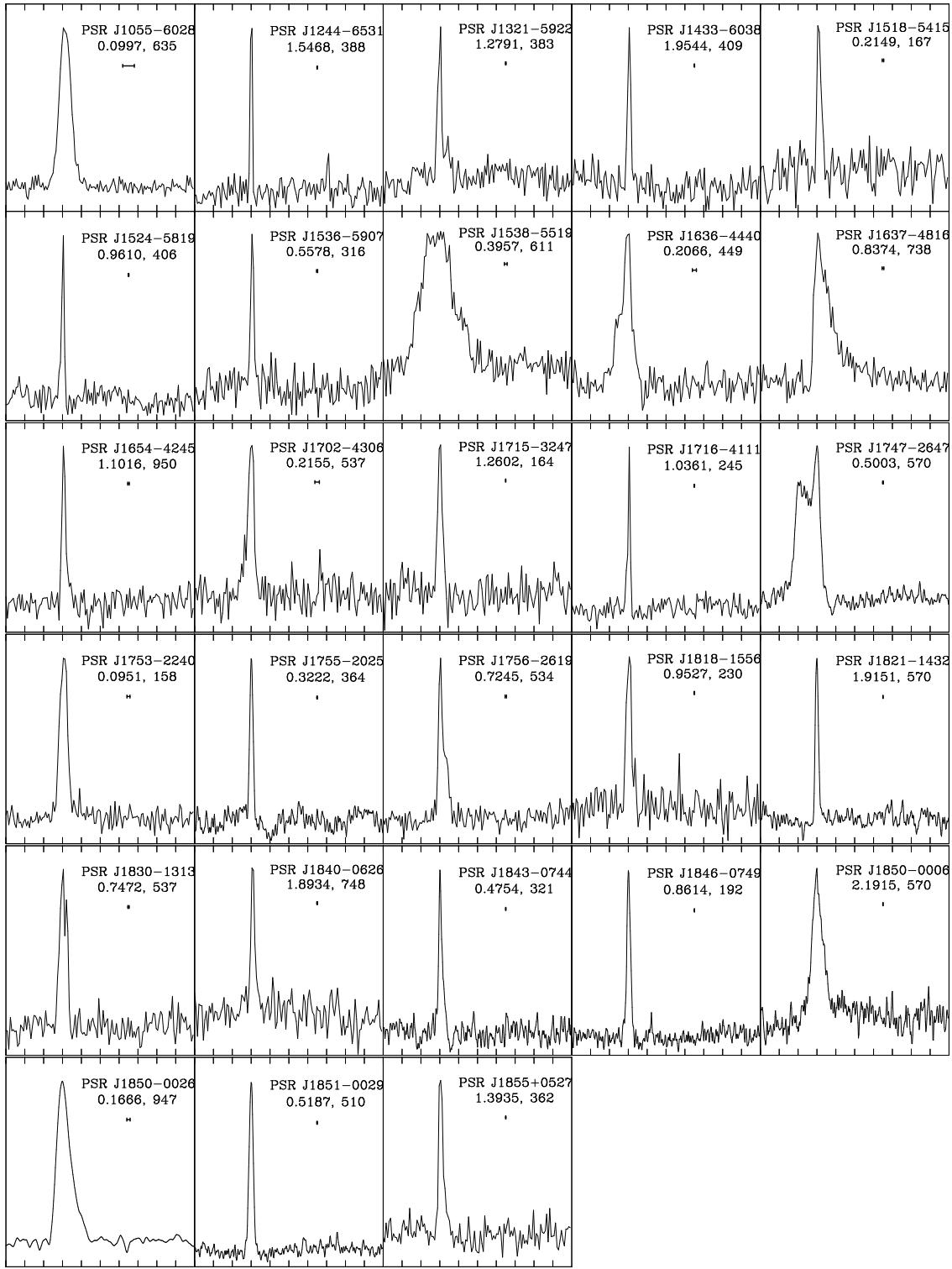}
\caption[Mean pulse profiles of each of the newly discovered pulsars.]{
\label{profiles}
Mean pulse profiles for each of the newly discovered pulsars measured
at 1374~MHz.  The profiles have been aligned such that the highest
peak in the profile is placed at a phase of 0.3.  Profiles have been
compensated for the effects of the high-pass filter in the survey
digitisation system \cite{mlc+01}.  Each plot covers the whole of the
pulsar period and is identified by the pulsar name, spin period (s)
and DM (${\rm cm}^{-3}{\rm pc}$).  The small
horizontal bar on each profile represents the effective time
resolution of the profile resulting from the effects of the
interstellar medium and instrumental broadening.

}
\end{figure*}

The details for the 28 new pulsars are presented in Tables
\ref{pos_tab} -- \ref{der_1}, and mean pulse profiles for each are
shown in Figure~\ref{profiles}.

Table \ref{pos_tab} shows the position of each source obtained by
timing observations, as well as the discovery signal-to-noise ratio,
which of the 13 beams of the multi-beam receiver the discovery was in
and radial distance from the centre of the beam in beam widths. This distance
may be greater than one if the flux density of the pulsar is variable or due to interference in the closet beam. In
addition, the measured flux density and pulse width is provided, as
measured from folded pulse profiles. Table \ref{prd_tab} shows the
basic parameters fitted in the timing solution for each of the
pulsars.  This includes the pulse period, period derivative, the epoch
of the period solution, the number of TOAs measured, the observation
time coverage, the RMS residual on the TOAs and the DM 
(DM) value (measured over a 288 MHz band centred on 1374 MHz).

Finally, Table \ref{der_1} details standard derived parameters for 
each new pulsar.  This includes the characteristic age, the surface
magnetic field strength, the rate of rotational energy loss and the
distance to the pulsar, as determined from the DM and two of the
standard electron density models of the Galaxy.  The DM distance is
also used to derive the luminosity and height above the Galactic plane
(z-distance).

\begin{table*}
\caption{
Positional information, flux densities and pulse widths for the 28
pulsars discovered in this work.  Beam denotes in which of the 13
beams of the multi-beam receiver the pulsar was discovered.  Radial
distance specifies the distance from the centre of the receiver beam
in which the pulsar was detected, as a fraction of the beam full-width half maximum.
Signal-to-noise-ratio values (S/N) are for the discovery detections.
S$_{1400}$ shows the mean flux density at 1400~MHz, with a nominal error of $30\%$. Pulse widths
are measured at $50\%$ of the peak pulse value.  Figures given in
parentheses are the 1-$\sigma$ error in the last-quoted digit.  }
\begin{center}
{\footnotesize \tt 
\begin{tabular}{lllrrrcrlrr}
\hline
PSR J & R.A. (J2000) & Dec. (J2000) & \multicolumn{1}{c}{$l$} & \multicolumn{1}{c}{$b$} & Beam & Radial & S/N & $S_{1400}$ & W$_{50}$  \\
 & (h~~~m~~~s) & (\degr ~~~\arcmin ~~~\arcsec) & \multicolumn{1}{c}{(\degr)} &\multicolumn{1}{c}{(\degr)} & & Distance & & (mJy) & (ms) \\
\hline
1055$-$6028 & 10:55:39.19(2) & $-$60:28:37.5(2) & 289.13 & $-$0.75 & 11 & 0.53 & 40.3 & 0.78 & 6\\
1244$-$6531 & 12:44:38.3(2) & $-$65:31:12(2) & 302.23 & $-$2.66 & 6 & 0.54 & 19.1 & 0.17 & 21\\
1321$-$5922 & 13:21:39.56(5) & $-$59:22:52.1(7) & 306.78 & 3.26 & 13 & 0.57 & 10.9 & 0.19 & 26\\
1433$-$6038 & 14:33:13.18(6) & $-$60:38:34.7(8) & 315.09 & $-$0.20 & 2 & 0.78 & 12.7 & 0.23 & 35\\
1518$-$5415 & 15:18:13.50(1) & $-$54:15:45.0(3) & 323.38 & 2.68 & 7 & 1.18 & 10.6 & 0.08 & 4\\
\\
1524$-$5819 & 15:24:24.65(2) & $-$58:19:14.1(4) & 321.91 & $-$1.20 & 1 & 0.80 & 9.9 & 0.09 & 13\\
1536$-$5907 & 15:36:17.72(2) & $-$59:07:03.6(6) & 322.72 & $-$2.73 & 1 & 0.60 & 12.4 & 0.22 & 13\\
1538$-$5519 & 15:38:40.83(9) & $-$55:19:46(2) & 325.22 & 0.13 & 4 & 1.16 & 14.4 & 0.42 & 86\\
1636$-$4440 & 16:36:16.53(2) & $-$44:40:25(1) & 339.18 & 1.80 & 6 & 0.68 & 13.4 & 0.38 & 10\\
1637$-$4816 & 16:37:58.68(4) & $-$48:16:12(2) & 336.71 & $-$0.83 & 6 & 0.28 & 27.0 & 0.74 & 67\\
\\
1654$-$4245 & 16:54:22.08(4) & $-$42:45:39(3) & 342.76 & 0.57 & 9 & 1.19 & 9.8 & 0.14 & 31\\
1702$-$4306 & 17:02:27.290(11) & $-$43:06:44(1) & 343.40 & $-$0.81 & 6 & 0.12 & 25.4 & 0.27 & 8\\
1715$-$3247 & 17:15:23.46(10) & $-$32:47:30(2) & 353.21 & 3.30 & 3 & 1.11 & 11.3 & 0.08 & 60\\
1716$-$4111 & 17:16:44.307(7) & $-$41:11:09.5(6) & 346.53 & $-$1.79 & 9 & 0.59 & 20.2 & 0.22 & 12\\
1747$-$2647 & 17:47:30.894(16) & $-$26:47:14(5) & 2.05 & 0.76 & 5 & 0.53 & 89.8 & 1.54 & 66\\
\\
1753$-$2240 & 17:53:39.830(5) & $-$22:40:52(8) & 6.30 & 1.66 & 10 & 0.51 & 11.9 & 0.15 & 3\\
1755$-$2025 & 17:55:35.801(6) & $-$20:25:00 & 8.48 & 2.42 & 2 & 0.69 & 14.7 & 0.18 & 6\\
1756$-$2619 & 17:57:19.39(2) & $-$26:19:08(9) & 3.57 & $-$0.89 & 2 & 1.00 & 9.5 & 0.10 & 18\\
1818$-$1556 & 18:18:51.95(2) & $-$15:56:04(2) & 15.09 & $-$0.23 & 7 & 0.26 & 14.0 & 0.49 & 26\\
1821$-$1432 & 18:21:39.777(15) & $-$14:32:53.2(19) & 16.63 & $-$0.17 & 13 & 0.65 & 11.0 & 0.22 & 37\\
\\
1830$-$1313 & 18:30:41.98(2) & $-$13:13:16.2(45) & 18.82 & $-$1.49 & 2 & 0.79 & 17.1 & 0.26 & 38\\
1840$-$0626 & 18:40:16.31(5) & $-$06:26:15.4(52) & 25.93 & $-$0.46 & 5 & 0.59 & 12.6 & 0.16 & 37\\
1843$-$0744 & 18:43:05.487(8) & $-$07:44:30.1(5) & 25.09 & $-$1.68 & 12 & 1.11 & 15.8 & 0.17 & 9\\
1846$-$0749 & 18:46:07.999(9) & $-$07:49:13.2(6) & 25.37 & $-$2.39 & 8 & 1.63 & 9.7 & 0.19 & 19\\
1850$-$0006 & 18:50:47.93(8) & $-$00:06:26.1(45) & 32.76 & 0.09 & 10 & 0.93 & 19.6 & 0.8 & 139\\
\\
1850$-$0026 & 18:50:14.714(4) & $-$00:26:11.6(2) & 32.41 & 0.07 & 13 & 1.94 & 19.3 & 1.77 & 13\\
1851$-$0029 & 18:51:55.093(10) & $-$00:29:58.1(5) & 32.54 & $-$0.33 & 9 & 0.72 & 29.1 & 0.44 & 13\\
1855$+$0527 & 18:55:15.073(18) & $+$05:27:40.7(9) & 38.23 & 1.64 & 10 & 0.70 & 15.3 & 0.24 & 37\\

\hline
\end{tabular}
}
\end{center}
\label{pos_tab}
\end{table*}

\begin{table*}
\caption{
Periods, period derivatives and DM for 28 pulsars
discovered in this work.  Also included is the MJD epoch for the
period value, MJD range of and the  number of TOAs used in the timing solution
and RMS deviation of the post-fit residuals.  Figures in parentheses are
the 1-$\sigma$ error in the last-quoted digit.  }
\begin{center}
{\footnotesize \tt
\begin{tabular}{llllllll}
\hline
PSR J & Period, $P$ & $\dot{P}$ & Epoch & N$_{\rm TOA}$ & Data Span & Residual & DM \\
      & (s)       & (10$^{-15}$) & (MJD) &         & (MJD)     & (ms)     & (cm$^{-3}$pc) \\
\hline
1055$-$6028 & 0.099660833480(3) & 29.5322(5) & 54362.2 & 44 & 54203-54649 & 0.6 & 635.9(2)\\ 
1244$-$6531 & 1.5468189400(3) & 7.18(10) & 54196.5 & 31 & 54029-54394 & 3.0 & 388(2)\\ 
1321$-$5922 & 1.27905759523(7) & 2.37(3) & 54212.0 & 28 & 54029-54442 & 3.1 & 383(2)\\ 
1433$-$6038 & 1.9544296353(5) & 5.63(4) & 54087.9 & 42 & 54029-54522 & 3.5 & 409(2)\\ 
1518$-$5415 & 0.214924800284(6) & 0.0406(14) & 54220.6 & 27 & 54047-54468 & 0.4 & 167.2(2)\\ 
\\
1524$-$5819 & 0.96104262694(10) & 126.016(10) & 54174.8 & 30 & 54047-54522 & 1.3 & 406.6(5)\\ 
1536$-$5907 & 0.55784066031(3) & 1.366(7) & 54211.6 & 33 & 54029-54468 & 1.8 & 316(2)\\ 
1538$-$5519 & 0.39573063893(8) & 0.041(15) & 54221.1 & 34 & 54047-54522 & 5.1 & 611(3)\\ 
1636$-$4440 & 0.20664850874(3) & 46.715(4) & 54411.5 & 35 & 54301-54649 & 1.3 & 449(1)\\ 
1637$-$4816 & 0.8373653018(3) & 5.834(15) & 54115.9 & 35 & 54029-54522 & 3.5 & 738(2)\\ 
\\
1654$-$4245 & 1.1015546927(7) & 51.15(3) & 54024.2 & 30 & 54029-54522 & 2.5 & 950(1)\\ 
1702$-$4306 & 0.215507334229(9) & 9.7852(10) & 54212.1 & 26 & 54029-54522 & 0.8 & 537(1)\\ 
1715$-$3247 & 1.26021405926(8) & 0.092(7) & 54244.6 & 14 & 54063-54646 & 4.9 & 164(3)\\ 
1716$-$4111 & 1.03606727254(3) & 2.882(5) & 54212.2 & 34 & 54029-54522 & 0.6 & 245.8(5)\\ 
1747$-$2647 & 0.500254454491(16) & 13.2412(18) & 54311.0 & 64 & 54034-54666 & 2.1 & 570(9)\\ 
\\
1753$-$2240 & 0.0951378101526(12) & 0.00079(14) & 54311.9 & 66 & 54028-54647 & 0.5 & 158.6(4)\\ 
1755$-$2025 & 0.322231178915(5) & 4.4217(6) & 54268.1 & 63 & 54038-54665 & 1.4 & 364.3(5)\\ 
1756$-$2619 & 0.72451374873(11) & 1.238(6) & 54223.2 & 28 & 54116-54647 & 1.4 & 534(2)\\ 
1818$-$1556 & 0.952708902877(19) & 0.7062(19) & 54263.4 & 73 & 54028-54675 & 1.7 & 230(4)\\ 
1821$-$1432 & 1.91513066964(8) & 5.373(8) & 54217.7 & 77 & 54036-54675 & 3.2 & 570(20)\\ 
\\
1830$-$1313 & 0.74718766820(8) & 1.714(17) & 54211.7 & 32 & 54028-54468 & 1.5 & 537(3)\\ 
1840$-$0626 & 1.8933526999(4) & 23.03(7) & 54211.2 & 25 & 54028-54522 & 4.4 & 748(8)\\ 
1843$-$0744 & 0.475392526561(10) & 13.3038(10) & 54274.1 & 65 & 54050-54675 & 0.8 & 321(5)\\ 
1846$-$0749 & 0.861379642641(15) & 5.187(3) & 54283.7 & 65 & 54069-54675 & 1.5 & 192(5)\\ 
1850$-$0006 & 2.1914979680(5) & 4.32(5) & 54274.6 & 51 & 54050-54665 & 15.2 & 570(20)\\ 
\\
1850$-$0026 & 0.1666339242514(20) & 39.1017(20) & 54263.7 & 82 & 54029-54666 & 0.7 & 947(5)\\ 
1851$-$0029 & 0.518721435025(9) & 4.7386(11) & 54143.7 & 129 & 53817-54666 & 1.9 & 510(20)\\ 
1855$+$0527 & 1.39348448168(8) & 267.207(8) & 54248.1 & 99 & 54028-54666 & 4.7 & 362(3)\\ 

\hline
\end{tabular}
}
\end{center}
\label{prd_tab}
\end{table*}

\begin{table*}
\caption{
Derived parameters for the 28 pulsars discovered in this
work. Included are: Characteristic age ($\tau_c$) in years;
characteristic surface dipole magnetic field strength ($B$) in Gauss;
the rate of rotational energy loss ($\dot{E}$) in erg s$^{-1}$; the
DM-derived distance using the \citeN{tc93} model ($D_{TC}$) and the
\citeN{cl02} model ($D_{CL}$) as well as the corresponding heights
above the Galactic plane ($z_{TC}$ and $z_{CL}$), all measured in kpc;
and finally the inferred 1374~MHz radio luminosity for the given
distance ($L_{TC}$ and $L_{CL}$) measured in mJy kpc$^2$.
The distance estimate, and therefore luminosity and $z$-height, may have large systematic
errors due to the electron density model and so care should be taken when using these values.}
\begin{center}
{\footnotesize \tt
\begin{tabular}{rrrrrrrrrr}
\hline
\multicolumn{1}{c}{PSR J} &
\multicolumn{1}{c}{$\log[\tau_c]$} &
\multicolumn{1}{c}{$\log[B]$} &
\multicolumn{1}{c}{$\log[\dot{E}]$} &
\multicolumn{1}{c}{$D_{\rm TC}$} &
\multicolumn{1}{c}{$D_{\rm CL}$} &
\multicolumn{1}{c}{$z_{\rm TC}$} &
\multicolumn{1}{c}{$z_{\rm CL}$} &
\multicolumn{1}{c}{$L_{\rm TC}$} &
\multicolumn{1}{c}{$L_{\rm CL}$} \\
 & & & & \multicolumn{2}{c}{(kpc)} & \multicolumn{2}{c}{(kpc)}&\multicolumn{2}{c}{(mJy kpc$^2$)} \\
\hline
J1055$-$6028 & 4.73 & 12.24 & 36.08 & 30.0 & 15.1 & $-$0.39 & $-$0.20 & 702.0 & 178.5\\ 
J1244$-$6531 & 6.53 & 12.53 & 31.89 & 30.0 & 9.4 & $-$1.39 & $-$0.44 & 153.0 & 15.0\\ 
J1321$-$5922 & 6.93 & 12.25 & 31.65 & 23.0 & 9.0 & 1.31 & 0.51 & 100.2 & 15.4\\ 
J1433$-$6038 & 6.74 & 12.53 & 31.48 & 9.9 & 6.3 & $-$0.03 & $-$0.02 & 22.8 & 9.1\\ 
J1518$-$5415 & 7.92 & 10.98 & 32.20 & 4.3 & 3.2 & 0.20 & 0.15 & 1.5 & 0.8\\ 
\\
J1524$-$5819 & 5.08 & 13.05 & 33.75 & 10.9 & 6.2 & $-$0.23 & $-$0.13 & 10.6 & 3.4\\ 
J1536$-$5907 & 6.81 & 11.95 & 32.49 & 12.0 & 6.0 & $-$0.57 & $-$0.29 & 31.5 & 7.9\\ 
J1538$-$5519 & 8.19 & 11.11 & 31.41 & 9.8 & 7.5 & 0.02 & 0.02 & 40.4 & 23.7\\ 
J1636$-$4440 & 4.85 & 12.50 & 35.32 & 9.5 & 6.6 & 0.30 & 0.21 & 34.4 & 16.4\\ 
J1637$-$4816 & 6.36 & 12.35 & 32.59 & 10.0 & 8.5 & $-$0.14 & $-$0.12 & 74.0 & 53.0\\ 
\\
J1654$-$4245 & 5.53 & 12.88 & 33.18 & 11.8 & 10.9 & 0.12 & 0.11 & 19.6 & 16.6\\ 
J1702$-$4306 & 5.54 & 12.17 & 34.59 & 7.1 & 6.7 & $-$0.10 & $-$0.09 & 13.8 & 12.0\\ 
J1715$-$3247 & 8.34 & 11.54 & 30.26 & 4.0 & 3.2 & 0.23 & 0.18 & 1.3 & 0.8\\ 
J1716$-$4111 & 6.76 & 12.24 & 32.00 & 4.8 & 4.0 & $-$0.15 & $-$0.12 & 5.0 & 3.5\\ 
J1747$-$2647 & 5.78 & 12.41 & 33.62 & 8.3 & 7.3 & 0.11 & 0.10 & 105.8 & 81.8\\ 
\\
J1753$-$2240 & 9.28 & 9.94 & 31.56 & 3.5 & 3.0 & 0.10 & 0.09 & 1.8 & 1.4\\ 
J1755$-$2025 & 6.06 & 12.08 & 33.72 & 8.7 & 6.4 & 0.37 & 0.27 & 13.6 & 7.4\\ 
J1756$-$2619 & 6.97 & 11.98 & 32.11 & 7.5 & 7.1 & $-$0.12 & $-$0.11 & 5.7 & 5.0\\ 
J1818$-$1556 & 7.33 & 11.92 & 31.51 & 4.1 & 3.9 & $-$0.02 & $-$0.02 & 8.1 & 7.6\\ 
J1821$-$1432 & 6.75 & 12.51 & 31.48 & 7.5 & 7.0 & $-$0.02 & $-$0.02 & 12.5 & 10.9\\ 
\\
J1830$-$1313 & 6.84 & 12.06 & 32.20 & 10.4 & 8.0 & $-$0.27 & $-$0.21 & 28.3 & 16.6\\ 
J1840$-$0626 & 6.11 & 12.82 & 32.11 & 9.2 & 8.9 & $-$0.07 & $-$0.07 & 13.5 & 12.6\\ 
J1843$-$0744 & 5.75 & 12.40 & 33.69 & 5.7 & 5.6 & $-$0.17 & $-$0.16 & 5.5 & 5.4\\ 
J1846$-$0749 & 6.42 & 12.33 & 32.51 & 4.4 & 3.9 & $-$0.18 & $-$0.16 & 3.7 & 2.9\\ 
J1850$-$0006 & 6.91 & 12.49 & 31.20 & 7.2 & 8.0 & 0.01 & 0.01 & 41.9 & 50.7\\ 
\\
J1850$-$0026 & 4.83 & 12.41 & 35.52 & 11.2 & 10.8 & 0.01 & 0.01 & 222.9 & 205.7\\ 
J1851$-$0029 & 6.24 & 12.20 & 33.11 & 6.9 & 7.6 & $-$0.04 & $-$0.04 & 20.8 & 25.2\\ 
J1855$+$0527 & 4.92 & 13.29 & 33.59 & 9.8 & 7.9 & 0.28 & 0.23 & 23.0 & 15.1\\ 

\hline
\end{tabular}
}
\end{center}
\label{der_1}
\end{table*}

\subsection{PSR J1753--2240: An eccentric binary system}
Initial confirmation observations of J1753--2240 showed that the
period was varying over time in a manner consistent with Doppler shifts caused by motion around an binary companion.
Analysis of observations during the
following month suggested that the pulsar was part of a binary system
with an orbital period of 13.6 days.  Since the observed period
variation was not sinusoidal we fit the observed periodicities with an
eccentric orbit using the {\sc fitorbit} software at Jodrell Bank.

Continued monitoring and timing of the
pulsar gives an accurate model of the orbit of the system, showing an
eccentricity of 0.3.  A full discussion of the detailed parameters of
this source is presented elsewhere \cite{kkl+08}.

\subsection{PSR J1850--0026: A young pulsar coincident with a supernova remnant}

\begin{figure}
\centering
\includegraphics[width=7.6cm]{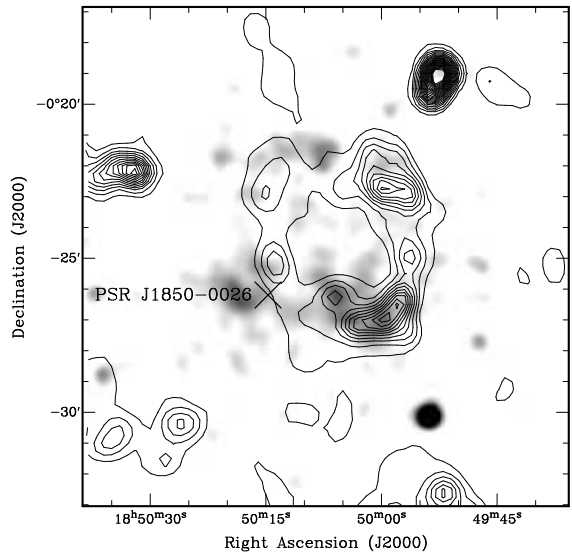}

\caption[Map of the position of PSR J1850--0026 in relation to SNR G32.45+0.1.]{
\label{1849map}
Map of the position of PSR J1850--0026 in relation to SNR G32.45+0.1,
showing XMM-Newton X-ray data in grey scale \cite{yuk+04} and a NVSS
1400-MHz radio map in contours \cite{ccg+98}.  The cross shows the position of the pulsar as derived from the timing solution.

}
\end{figure}

PSR~J1850-0026 is a young pulsar with a characteristic age of 68~kyr
and is spatially coincident with SNR G32.45+0.1, listed in the Green
supernova catalogue \cite{gre06}. \citeN{yuk+04} present X-ray observations of this
source, which they suggest indicate a shell-like structure coincident with the position of
a radio shell in the NVSS 1400-MHz VLA survey \cite{ccg+98}.  The
shell structure is somewhat irregular, but it has an average angular radius of
around 2 to 3 arc minutes.  The position of the pulsar is well defined
by timing measurements, and is $\sim2.5$ arc minutes south west of the
nominal centre of the remnant.  This position is shown in relation to
the remnant in Figure \ref{1849map}.
There are no known gamma-ray or TeV sources close to the position of PSR~J1850-0026.

By considering a random sample of positions in this area of the sky,
one can estimate the probability of a chance alignment between PSR J1850-0026 and SNR G32.45+0.1.
We chose a large sample of randomly distributed sky positions with 
$-50^\circ < l < 50^\circ$ and $|b| < 1^\circ$ and searched for coincident
positions in the catalogue of supernova remnants.
At a separation of 2.5 arc minutes, the probability of a chance alignment is
approximately $0.25\%$.
Given that there are 218 known pulsars in this area of the sky, the expected
number of chance alignments is $0.7$.
If a reduced set of only the 28 known pulsars with ages less than 100 kyr are selected,
then only $0.07$ chance alignments are expected.

Hydrogen absorption measurements suggest that the distance to the
remnant is $17~\pm~11$~kpc \cite{yuk+04}, which is compatible with the
estimated distance to the pulsar of $11\pm5$~kpc, calculated using
the measured DM and the NE2001 \cite{cl02} electron
density model.

Reliable association of pulsars with supernova remnants is difficult,
and it is impossible to completely rule out the possibility of a
chance alignment.  There are however several tests that can give a good
indication that an association is likely.  The most significant of these is a measurement of the
proper motion of the pulsar.  It would be expected that the pulsar
would originate in the centre of the remnant, so that the proper
motion should be directed away from this point.  Measurement of proper
motion usually requires several years of observation, and accurate
measurements have been made on other similar pulsars
\cite{hlk+04}.
Assuming that the pulsar was born in the supernova explosion,
we can use the pulsar's characteristic age (but see \citealp{kbm+03} for potential problems)
and the projected distance of the pulsar from the centre of the explosion (from the distance
estimates and the angular separation), we can estimate the velocity of
the pulsar perpendicular to the line of sight.
The distribution of pulsar velocities is well known \cite{hllk05}, and therefore if the
computed velocity lies outside of the range of measured velocities
then association is unlikely.  Unfortunately there are large
systematic errors in both the distance and the age measurements, such
that the velocity measurement is highly uncertain.  For
PSR~J1850--0026, the expected transverse velocity is approximately $140$ km/s
for a nominal distance of 11 kpc, age of 68 kyr and a separation of 2.5
arc minutes.  This velocity scales proportionally to changes in any of
the three input parameters, although the computed value is well within
the range of measured pulsar velocities.
Given the above, it appears that there is a significant probability
that PSR J1850--0026 is associated with SNR G32.45+0.1.

\section{Statistical analysis}
\label{stats}
In order to determine if the new candidate selection techniques are
selecting a statistically different set of pulsars, we compare our
discoveries with those presented in \citeN{fsk+04}.
We choose this sample of 142 pulsars because the survey parameters and 
processing code used were identical, thus the only difference is in
the means of filtering the millions of candidates selected by the search algorithms.
Unfortunately due to the fact that {\sc JReaper} was designed to automatically detect and
remove known pulsars from the results, it is only possible to compare the discovered pulsars,
rather than all detected pulsars.
An attempt to consider the effectiveness of {\sc JReaper} at detecting pulsars is
presented in Section \ref{eff}.

The distribution of pulsar periods is split into two populations, one
comprising millisecond pulsars and the other normal pulsars.  The set
of newly discovered pulsars contains no very short-period
pulsars, i.e those with a spin period of less than 20~ms. Extrapolation of the expected number of short-period pulsars
from the previous analysis suggests that the new results should
contain $2.5\pm 1.6$ short-period pulsars.  The small number of new pulsar discoveries  make it
difficult to determine if the absence of new discoveries of short-period pulsars is really
significant.

\begin{figure}\centering
\includegraphics[width=7cm]{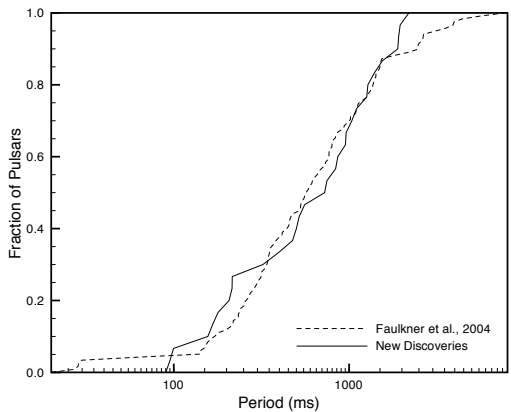}
\caption[Cumulative distribution plot of periods for two samples of pulsars.]{
\label{pmj_per_cumu}
Cumulative distribution plot of periods for two samples of pulsars,
using a logarithmic scale.  }
\end{figure}
\begin{figure}\centering
\includegraphics[width=7cm]{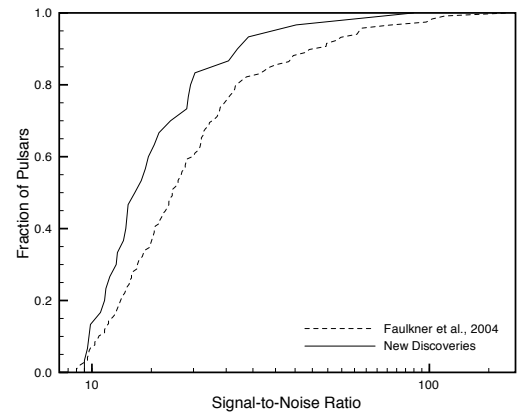}
\caption[Cumulative distribution plot of SNRs for two samples of pulsars.]{
\label{pmj_snr_cumu}
Cumulative distribution plot of signal-to-noise ratios for two samples
of pulsars, using a logarithmic scale.  }
\end{figure}

\begin{figure}\centering
\includegraphics[width=7cm]{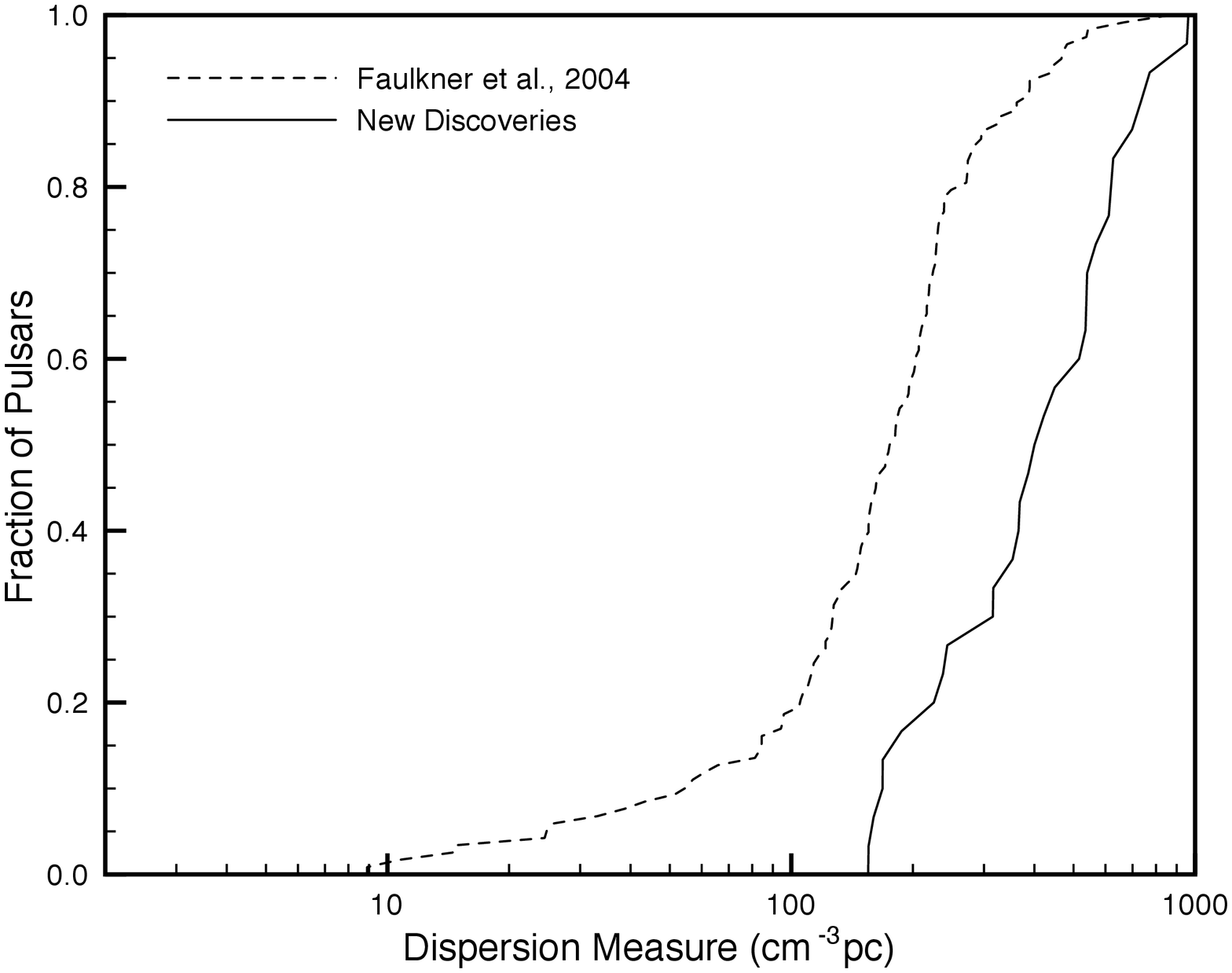}
\caption[Cumulative distribution plot of DMs for two samples of pulsars.]{
\label{pmj_dm_cumu}
Cumulative distribution plot of DMs for two samples of
pulsars, using a logarithmic scale.  }
\end{figure}

Figure \ref{pmj_per_cumu} shows the cumulative distribution of the
rotational periods of the two samples.  A two-sample Kolmogorov-Smirnov
test (KS test) determines that these samples are statistically
indistinguishable.  This implies that there are no difference in the period distributions of
pulsars found by the two searches.

Figure \ref{pmj_snr_cumu} again shows a cumulative distribution plot,
this time showing the signal-to-noise ratios for the detections in the
two data sets.  
Although only significant at the 95\% level, it appears that the new search methods may improve sensitivity to low
signal-to-noise-ratio candidates.  We can speculate that this difference may
be due to the fact that brighter sources are more likely to have been
detected in earlier, predominantly signal-to-noise-ratio based searches.

\begin{figure}\centering
\includegraphics[height=7cm,angle=-90]{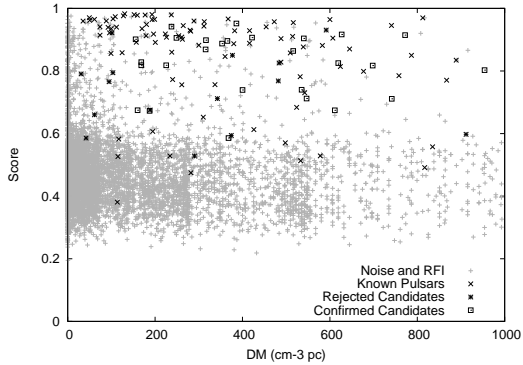}
\caption[DM and score values for a selection of candidate pulsars]{
\label{dm_score}
The DM and score values for a sample of pulsar candidates from the PMPS.
The darker points show the discoveries presented here, the known pulsars and candidates that were selected for re-observation but were not confirmed.
The lighter grey points are candidates produced by the search processing that were not selected for re-observation and are likely due to noise or interference.
The vertical bands visible in the noise candidates are caused by the re-sampling of the data at higher DM values.
}
\end{figure}

Figure \ref{pmj_dm_cumu} shows a similar plot, displaying the
cumulative distribution of the DMs of the two samples.  In
this case it is clear that the distributions differ, and the
KS test verifies that these distributions differ with a significance
level of greater than $99\%$.  The mean DM for the
Faulkner et al. pulsars is 199 cm$^{-3}$pc, considerably lower than the mean value
of 453 cm$^{-3}$pc in the new data set.  
Clearly it is important to determine if the new techniques are selecting against low-DM pulsars or if it is picking up high-DM pulsars that have previously been missed.
Figure \ref{dm_score} indicates that the score of pulsars does not appear to be a function of the DM value, and although there are a greater number of false positives at lower DM values, the ratio of false positives (score $>$ 0.7) to number of candidates produced remains constant over the entire DM range.
Therefore, at least in terms of the scoring algorithms, there appears to be no selection effect against low-DM pulsars.


It seems then that the pulsars presented in this work are from the
same distribution of pulse periods as sampled in the earlier work,
although we have tended to discover pulsars with a lower flux density and
a greater DM.

It is also worth noting that the two new detections with the highest
signal-to-noise ratio have periodicities of 99.7 ms and 500.3 ms,
i.e. very close to the 10th and 2nd harmonics of 1 second, often a
strong interference signal.  It is probable therefore that these two
sources were not detected in the original processing because they were
close to interference signals and therefore ignored.  This suggests
that whilst the new techniques are best suited for detecting weaker
pulsars in the data, they are also effective at distinguishing
stronger pulsars from interference signals.

\subsection{Effectiveness of JReaper as a search tool}
\label{eff}
To assess the effectiveness of a tool such as {\sc JReaper} is
difficult because the final choice of candidates to re-observe is still
dependant on the user.

\begin{figure}
\includegraphics[height=7cm,angle=-90]{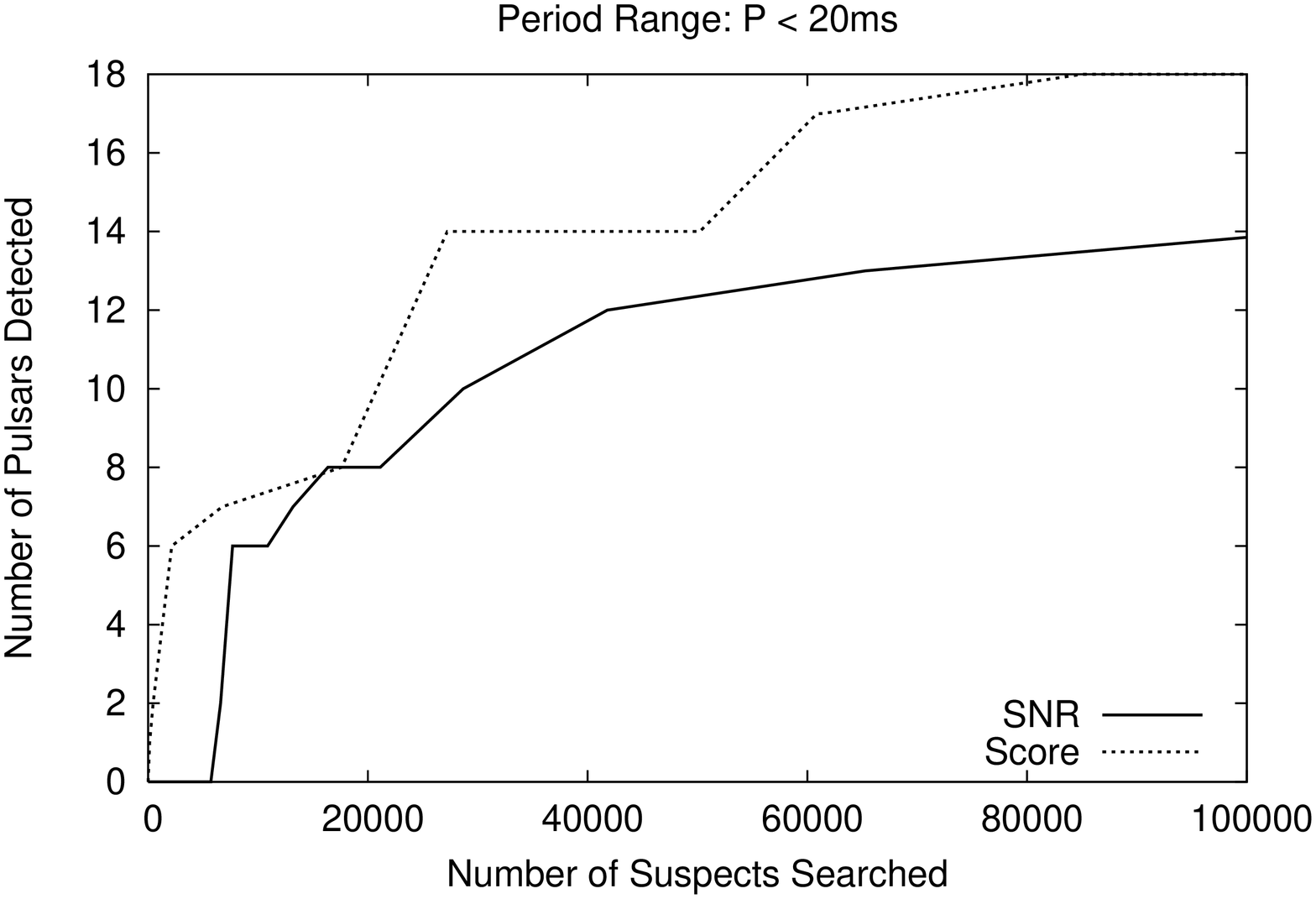} \\
\includegraphics[height=7cm,angle=-90]{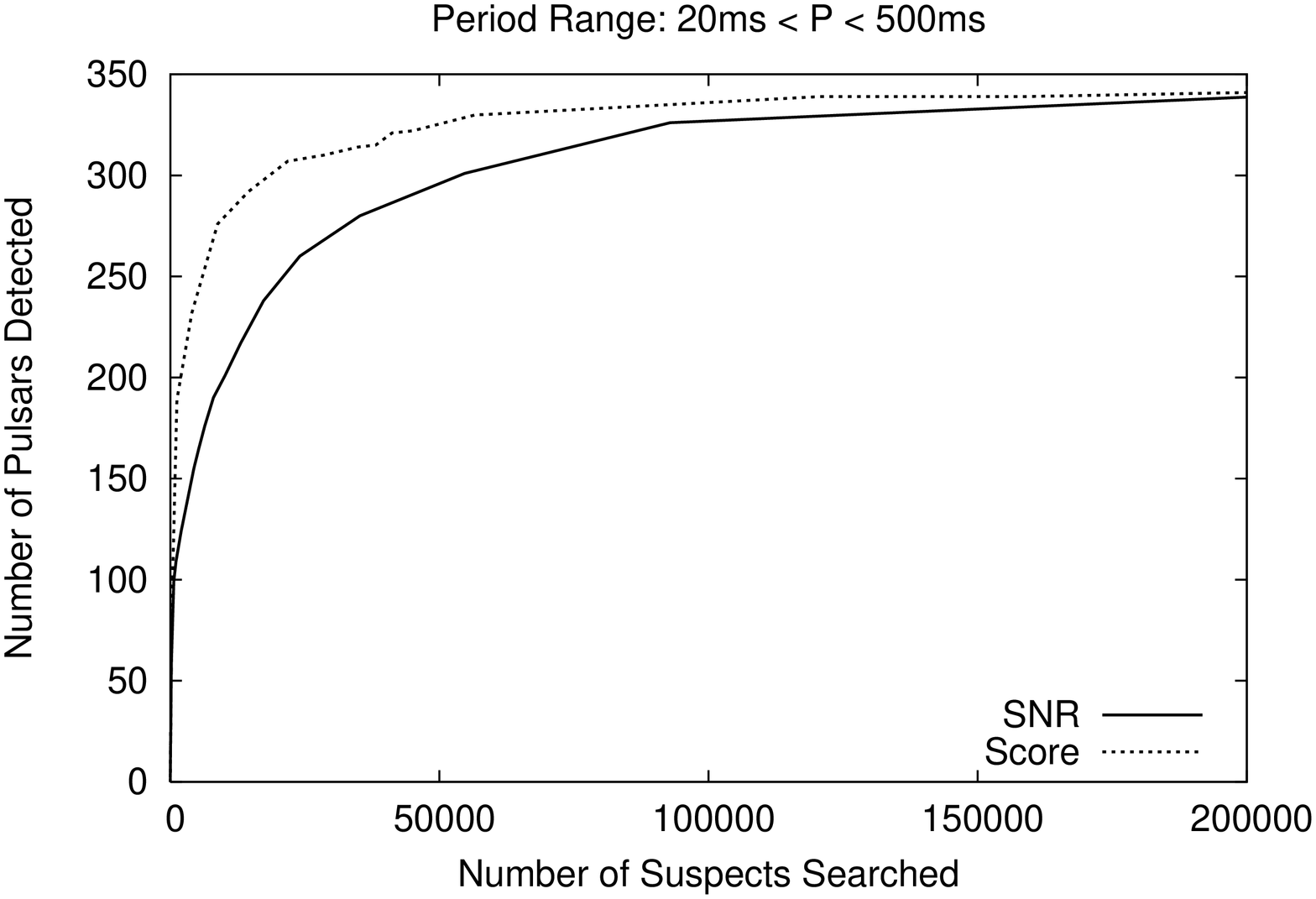} \\
\includegraphics[height=7cm,angle=-90]{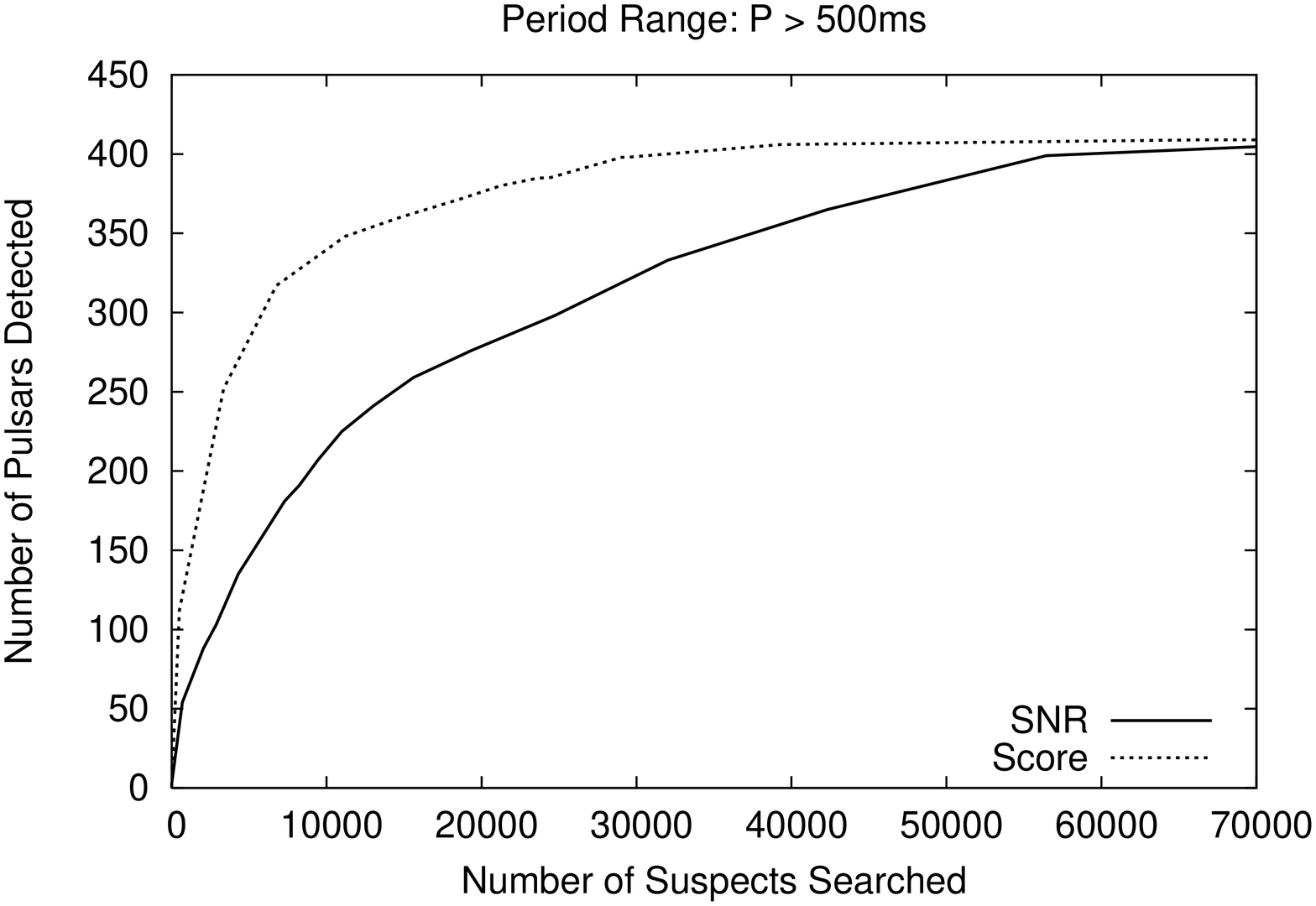}  \\
\caption[Comparison of the effectiveness of thresholding by SNR and Score]{
\label{scorecomparison}
These figures show the number of pulsars detected per candidate
searched when analysing candidates ordered by signal-to-noise ratio or
{\sc JReaper} score, for candidates in three ranges of period.  }

\end{figure}

The most significant, and testable, enhancement provided by {\sc
JReaper} is the pulsar scoring system.  A simple test of the score
system's effectiveness can be carried out by performing a blind search
for known pulsars.  This can be implemented by sorting all candidates
by their score and determining how many pulsars are found per
candidate searched.  This can be compared to the same process, but
sorting candidates by the more standard signal-to-noise ratio.  The
candidates were separated into three period ranges; less than 20ms,
between 20ms and 500ms and greater than 500ms.  The results of this
analysis are presented in Figure \ref{scorecomparison}.

This analysis clearly shows that for all period ranges the pulsar
scoring system is effective at selecting pulsars from other
candidates, when compared to the method based purely on the 
signal-to-noise ratio.  Unfortunately the relative scarcity of
millisecond pulsars makes the difference difficult to determine,
although each pulsar is detected earlier when ordered by score than by
signal-to-noise ratio.  The most significant effect is for the period
range greater than 500ms.  This is most likely to be due to the fact
that for longer periods there is an abundance of impulsive
interference, which the scoring can easily distinguish from pulsar-like
signals.  In this region, to detect $90\%$ of the detectable
known pulsars, one must search $\sim 4$ times fewer candidates compared to
a ranking based simply on signal-to-noise ratio.  For the 20-500ms region,
this figure drops to $\sim 3$ times.

Clearly this blind search is not the most efficient means of searching
for pulsars.  One would need to individually inspect of the order of $10^5$
candidates to find $90\%$ of the $\sim 700$ known pulsars in the data.
Whilst this may be possible, it would be more efficient to use the
various {\sc JReaper} plots to more intelligently select the
candidates to view.  The blind search does however illustrate the relative effectiveness
of the scoring system over an approach based upon the signal-to-noise ratio.

\section{Conclusion}
\label{conclusions}
Using new candidate selection techniques we discovered 28 pulsars that
were initially missed in the analysis of the Parkes Multi-beam Pulsar
Survey.  Amongst these discoveries are an eccentric binary system and
a young pulsar spatially coincident with a supernova remnant.
Although the fractional increase in discoveries is relatively small,
application of these new techniques comes with little cost.  We
believe that future search efforts will benefit from application of
candidate selection methods as presented here, in addition to the
established methods.

\section*{Acknowledgements}
This research was partly funded by grants from the Science \& Technology Facilities Council, UK. The Australia Telescope
is funded by the Commonwealth of Australia for operation as a National
Facility managed by the CSIRO.
The authors would like to thank the following people for their contributions to observing at Parkes:
M. Burgay, A. Noutsos, G.Hobbs, J. O'Brien, I. Stairs, A. Corongiu, M. Purver, R. Smits and C. Espinoza.

\bibliographystyle{mnras}
\bibliography{journals,myrefs,modrefs,psrrefs,crossrefs} 


\end{document}